\documentclass[letter, longauth]{aa} 

\usepackage{graphicx}
\usepackage[breaklinks=true]{hyperref} 
\hypersetup{colorlinks=true,citecolor=blue}

\begin{document} 
\title{Variability and dust filtration in the transition disk J160421.7-213028 observed in optical scattered light \thanks{Based on observations performed with VLT/SPHERE under program ID 095.C-0693(A)}}
   
\author{P.~Pinilla\inst{1}, J.~de Boer\inst{1}, M.~Benisty \inst{2}, A.~Juh\'asz \inst{3}, M.~de~Juan~Ovelar\inst{4}, C.~Dominik\inst{5}, H.~Avenhaus\inst{6}, T.~Birnstiel\inst{7}, J.~H.~Girard\inst{8}, N.~Huelamo\inst{9}, A.~Isella\inst{10}, and J.~Milli\inst{8}}
\institute{Leiden Observatory, Leiden University, P.O. Box 9513, 2300RA Leiden, The Netherlands\\
\email{pinilla@strw.leidenuniv.nl}
\and Univ. Grenoble Alpes, IPAG, F-38000 Grenoble, France CNRS, IPAG, F-38000 Grenoble, France
\and Institute of Astronomy, Madingley Road, Cambridge CB3 OHA, United Kingdom
\and Astrophysics Research Institute, Liverpool John Moores University,146 Brownlow Hill, Liverpool L3 5RF, UK
\and Astronomical Institute Anton Pannekoek, University of Amsterdam, PO Box 94249, 1090 GE Amsterdam, The Netherlands
\and Departamento de Astronom\'ia, Universidad de Chile, Casilla 36-D, Santiago, Chile
\and Harvard-Smithsonian Center for Astrophysics, 60 Garden Street, Cambridge, MA 02138, USA
\and European Southern Observatory (ESO), Alonso de C\'ordova 3107, Vitacura, Casilla 19001, Santiago, Chile
\and Centro de Astrobiolog\'ia (INTA-CSIC); ESAC Campus, PO Box 78, 28691 Villanueva de la Canada, Spain
\and Department of Physics \& Astronomy, Rice University, 6100 Main Street, Houston, TX 77005, USA}

  \abstract
   {Protoplanetary disks around young stars are the birth-sites of planets. Spectral energy distributions and direct images of a subset of disks known as transition disks reveal dust-depleted inner cavities. Some of these disks show asymmetric structures in thermal submillimetre emission and optical scattered light.  These  structures can be the result of planet(s) or companions embedded in the disk.}
   {We aim to detect and analyse the scattered light  of the transition disk J160421.7-213028, identify disk structures, and compare the results with previous observations of this disk at other wavelengths.}
   {We obtained and analysed new polarised intensity observations of the transition disk J160421.7-213028  with VLT/SPHERE using the visible light instrument ZIMPOL at $R'$-band (0.626\,$\mu$m). We probed the disk gap down to a radius of confidence of 0.1'' (${\sim}15$~AU at 145~pc). We interpret the results in the context of dust evolution when planets interact with the parental disk.}
   {We observe a gap from 0.1 to 0.3'' (${\sim}15$ to 40~AU) and a bright annulus as previously detected by HiCIAO $H$-band observations at $1.65\mu$m. The radial width of the annulus is around $40~$AU, and its centre is at ${\sim}61~$AU from the central star.  The peak of the reflected light at  0.626\,$\mu$m is located 20~AU inward of the cavity detected in the submillimetre. In addition, we detect a dip at a position angle of ${\sim}46.2 \pm 5.4^\circ$. A dip was also detected with HiCIAO, but located at ${\sim}85^\circ$. If the dip observed with HiCIAO is the same, this suggests an average dip rotation of ${\sim}12^\circ/$year, which is inconsistent with the local Keplerian angular velocity of $\sim$0.8$^\circ$/yr at $\sim$61~AU.}
   {The spatial discrepancy in the radial emission in J160421.7-213028 at different wavelengths is consistent with dust filtration at the outer edge of a gap carved by a massive planet. The dip rotation can be interpreted as fast variability of the inner disk and/or the presence of a warp or circumplanetary material of a planet at ${\sim}9.6$~AU.}

   \authorrunning{P.~Pinilla}
   \titlerunning{SPHERE observations of J1604}
   \maketitle
%

\section{Introduction} \label{sec:intro}
Recent observations of transition disks (TDs) have provided insight into the processes of planet formation and circumstellar disk dissipation \citep[e.g.][]{espaillat2014}. High-contrast imaging in the optical and near-infrared regime and observations at millimetre wavelengths not only revealed large clear inner cavities, but also several types of structures such as spiral arms, asymmetries, dips,  and disk eccentricities \citep[e.g.][]{garufi2013, marel2013, quanz2013, avenhaus2014, thalmann2014, benisty2015}. Different processes that are not mutually exclusive, can rule the disk evolution and create the observed structures, such as photoevaporation or magneto-rotational instabilities \citep[e.g.][]{alexander2014, flock2015}.

Spatially resolved observations at different wavelengths are  required to distinguish the physical fingerprints that  each of these mechanisms leaves on the dust and gas distribution of protoplanetary disks. For example, the  spatial segregation between small and large particles, as observed for several TDs \citep[e.g.][]{garufi2013}, is a natural consequence of  filtration effects caused by particle traps \citep[e.g.][]{rice2006, zhu2012, dejuanovelar2013}. One way to form a particle trap is planet-disk interaction: at the outer edge of a planetary gap, a region with positive pressure gradient can stop the fast inward migration of large dust particles \citep[e.g.][]{pinilla2012, pinilla2015}.

   \begin{figure*}
   \centering
   \includegraphics[width=17cm]{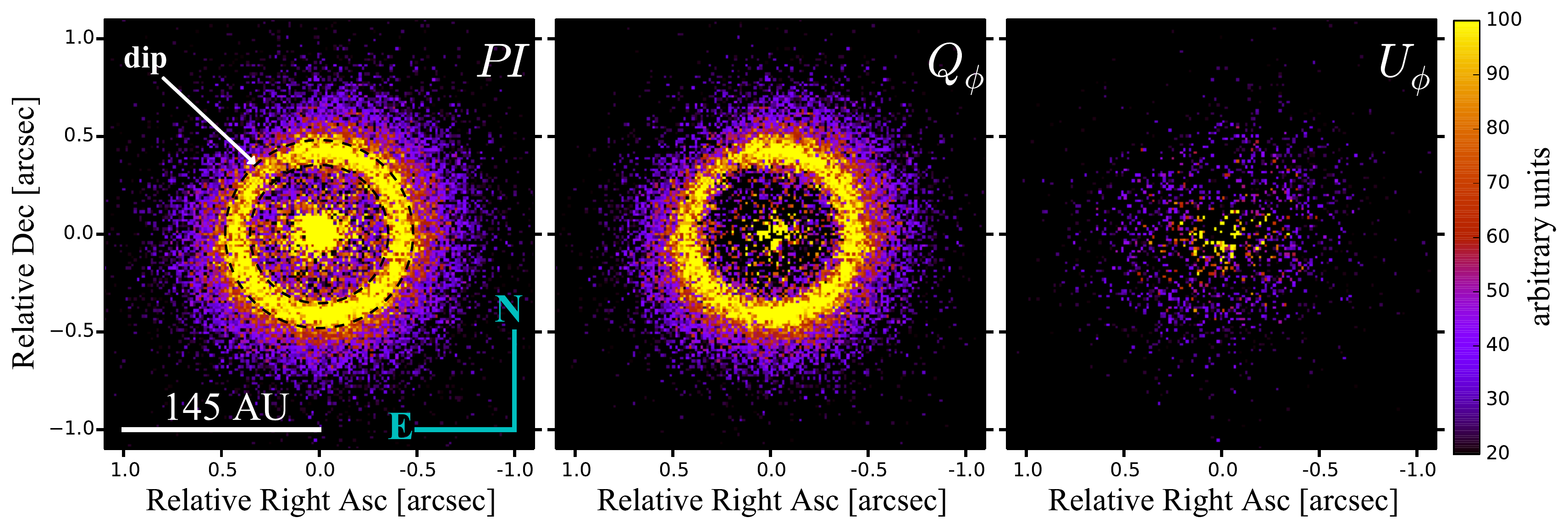}
   \caption{{\small $R'$ band (0.626\,$\mu$m) VLT/SPHERE/ZIMPOL images of J1604 (they are not scaled by $r^2$). From left to right: polarised intensity ($PI$),  polar-coordinate Stokes parameters $Q_\phi$ and $U_\phi$ respectively, such that $PI=\sqrt{Q_\phi^2+U_\phi^2}$. The clean $U_{\phi}$ image shows that we had an optimal correction for the instrumental polarisation. The colour scale is the same for the three panels; it is linear and in arbitrary units. The dashed lines in the left panel correspond to 0.35 and 0.48~arcsec,  which is the region where the azimuthal profile is calculated in Fig.~\ref{azimuthal_variation} to distinguish the dip.} \vspace{-0.2cm}}
   \label{sphere_images}
    \end{figure*}

In this letter, we present polarimetric differential imaging (PDI) of the transition disk around J160421.7-213028 (hereafter J1604), obtained with the subsystem ZIMPOL of the SPHERE instrument of the Very Large Telescope (VLT), at $R'$ band (0.626\,$\mu$m). This disk is a member of the Upper Scorpius association \citep{preibisch1999}, which is 5-10~Myr old \citep{pecaut2012}, and it is located at ${\sim}145$~pc \citep{zeeuw1999}. The disk is an excellent candidate to identify structures because it has one of the largest cavities reported in TDs  and is seen almost face-on \citep[${\sim}6^\circ$,][]{mathews2012}. Its cavity was resolved with observations from the Submillimeter Array (SMA) \citep{mathews2012} and was recently observed with the Atacama Large Millimeter/submillimeter Array (ALMA) in Cycle~0 \citep{zhang2014}, with a beam size of $0.73''\times0.46''$ ($106\times67$~AU at 145 pc). The observations with ALMA showed that the gas cavity is much smaller than the mm-dust cavity \citep[radius of 31~AU inferred from CO emission vs. 79~AU from the continuum,][]{zhang2014}.  In addition, near-infrared polarised intensity images obtained with HiCIAO at $1.6~\mu$m  \citep{mayama2012} revealed an asymmetric ring of ${\sim}$63~AU radius, with a dip located at a position angle (P.A., measured from north to east) of $85^{\circ}$. A tentative second dip was suggested at P.A. of $255^{\circ}$.

This letter is organised as follows. In Sect.~\ref{sec:observations} we describe the observations and data reduction.  The main results from the data analysis and the comparison with previous observations of this disk is presented in Sect.~\ref{sec:results}. We conclude with the discussion and perspectives in Sect.~\ref{sec:discussion}.

\section{Observations and data reduction} \label{sec:observations}

VLT/SPHERE/ZIMPOL observations of J1604 were performed on June~10, 2015, as part of the observing run 095.C-0693(A). 
We have used field tracking, polarimetric (P2) mode with the $R'$ filter 
($\lambda_0 = 0.626\,\mu$m, FWHM $= 0.148\,\mu$m) for both cameras.
Although there is currently no alternative to ZIMPOL for polarimetric imaging of southern targets in the visible, the R $= 11.8$ magnitude of J1604 \citep{cutri2003} poses a serious challenge for SAXO, the 
SPHERE extreme adaptive optics `xAO' \citep{beuzit2006, fusco2014}.
A beamsplitter divides the visible light of the star between ZIMPOL and the wave front sensor (WFS) of SAXO.
Observing in $R'$ band allowed us to use the dichroic beamsplitter, which sends all visible light except for the $R$ band to the WFS, thus ensuring an optimal AO correction.
During the observations, the seeing conditions were moderate to poor (0.9'' - 1.2''), which caused the Strehl ratio to vary by more than a factor of two. 
The median Strehl ratio obtained was ${\sim} 3.5\,\%$, resulting in a FWHM of ${\sim} 53 \times 47$~mas. The observing block was divided into six cycles of the half-wave plate (HWP), during which the HWP moved to four angles ($\theta_{\rm{hwp}} = 0^\circ; 45^\circ; 22.5^\circ$; and $67.5^\circ$) to measure the two linear Stokes components. 
For each HWP position, two exposures were taken of 120~s each, which adds up to 96 minutes of 
total observing time.

The data reduction is described in detail by De Boer et al. in prep., based on the description of ZIMPOL
by \cite{schmid2012}.
The pixels of the two detectors have a plate scale of $3.5885 \pm 0.0025~$mas per pixel (Ginski et al. in prep.).
We binned the pixels to a size of $14.354~$mas.
We then substracted the two different states of the ferro-electric liquid crystal (FLC),
the $0$ and $\pi$ frames \citep{schmid2012}, the ordinary and extra-ordinary beams of the polarising beam splitter; and the two matching HWP angles to obtain
Stokes $Q$ (for $\theta_{\rm{hwp}} = 0^\circ$ and $45^\circ$) and $U$ (for $\theta_{\rm{hwp}} = 22.5^\circ$ and $67.5^\circ$).

Figure~\ref{sphere_images} shows the polarised intensity $PI$ image and the polar-coordinate Stokes parameters $Q_{\phi}$ and $U_{\phi}$ \citep{schmid2006}, computed according to
\begin{eqnarray}
	PI           &=& \sqrt{Q^2 + U^2},\\
	Q_{\phi} &=&  Q \times \cos{2\phi} +  U \times \sin{2\phi}, \\
        U_{\phi} &=&   Q \times \sin{2\phi} -  U \times \cos{2\phi},
\end{eqnarray}

\noindent where $\phi$ is the position angle. 

By measuring the signal over an unpolarised region surrounding the star in the $Q$ and $U$ images, 
we determined the instrumental polarisation (IP), for which we corrected using the method described by \cite{canovas2011}.

The models of \cite{canovas2015} show that it is possible 
for an astrophysical signal to appear in the $U_\phi$ images, even when single-scattering dominates. 
However, this $U_\phi$ component only occurs for disks at high inclination ($i> 40^\circ$).
Since the disk of J1604 has an inclination of  $i= 6 \pm 1.5^\circ$ \citep{mathews2012}, 
we can use the assumption that the polarised scattered light is entirely tangential 
and therefore only appear in $Q_\phi$, while $U_\phi$ should
not contain any scattered light signal from the disk. We optimised our IP correction by minimising the
$U_\phi$ signal and found an optimum when we used an annulus of $10 \le r \le 15$ binned pixels.

   \begin{figure*}
   \centering
   \begin{tabular}{cc}  
   \includegraphics[width=6.5cm]{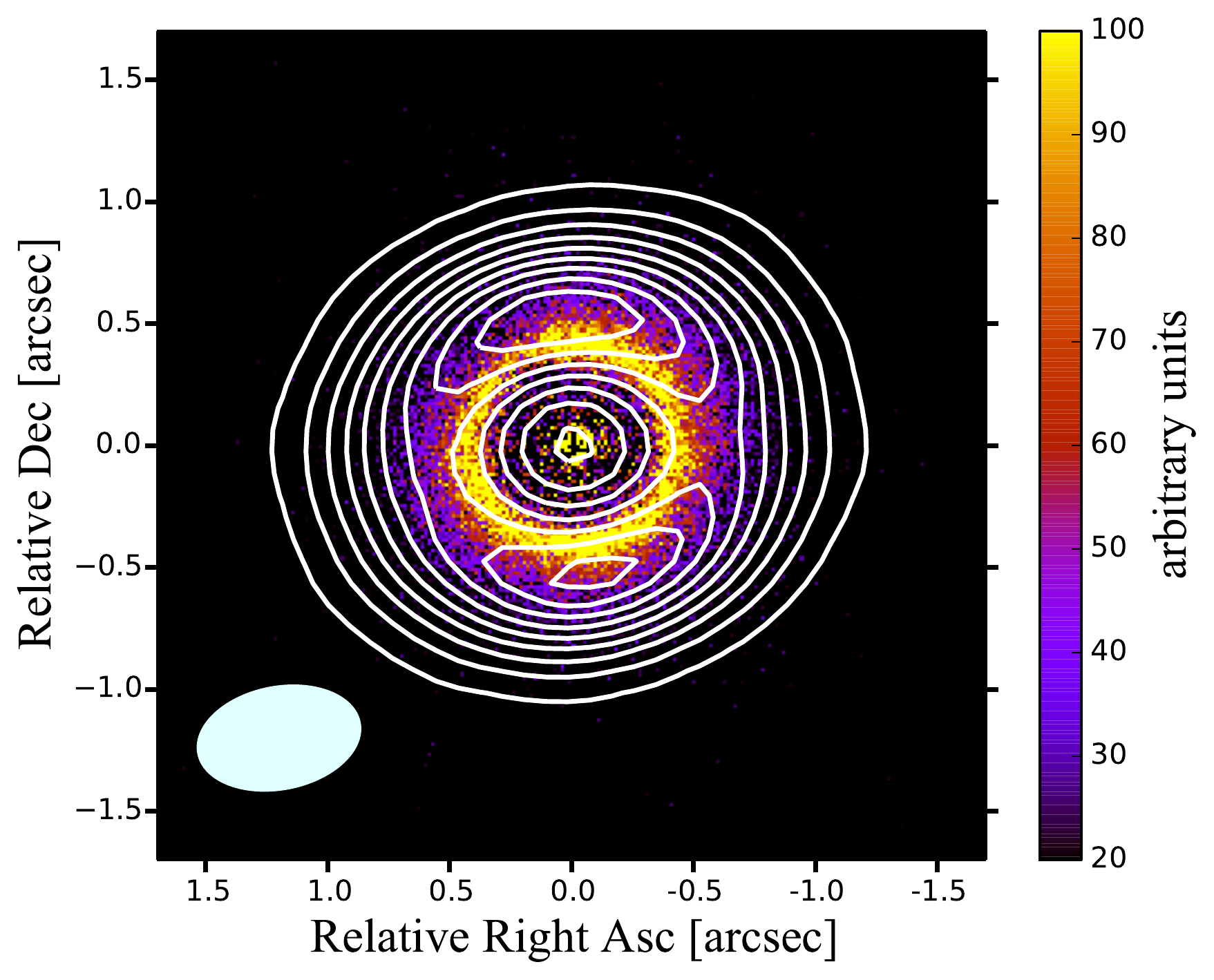}&\includegraphics[width=8.0cm]{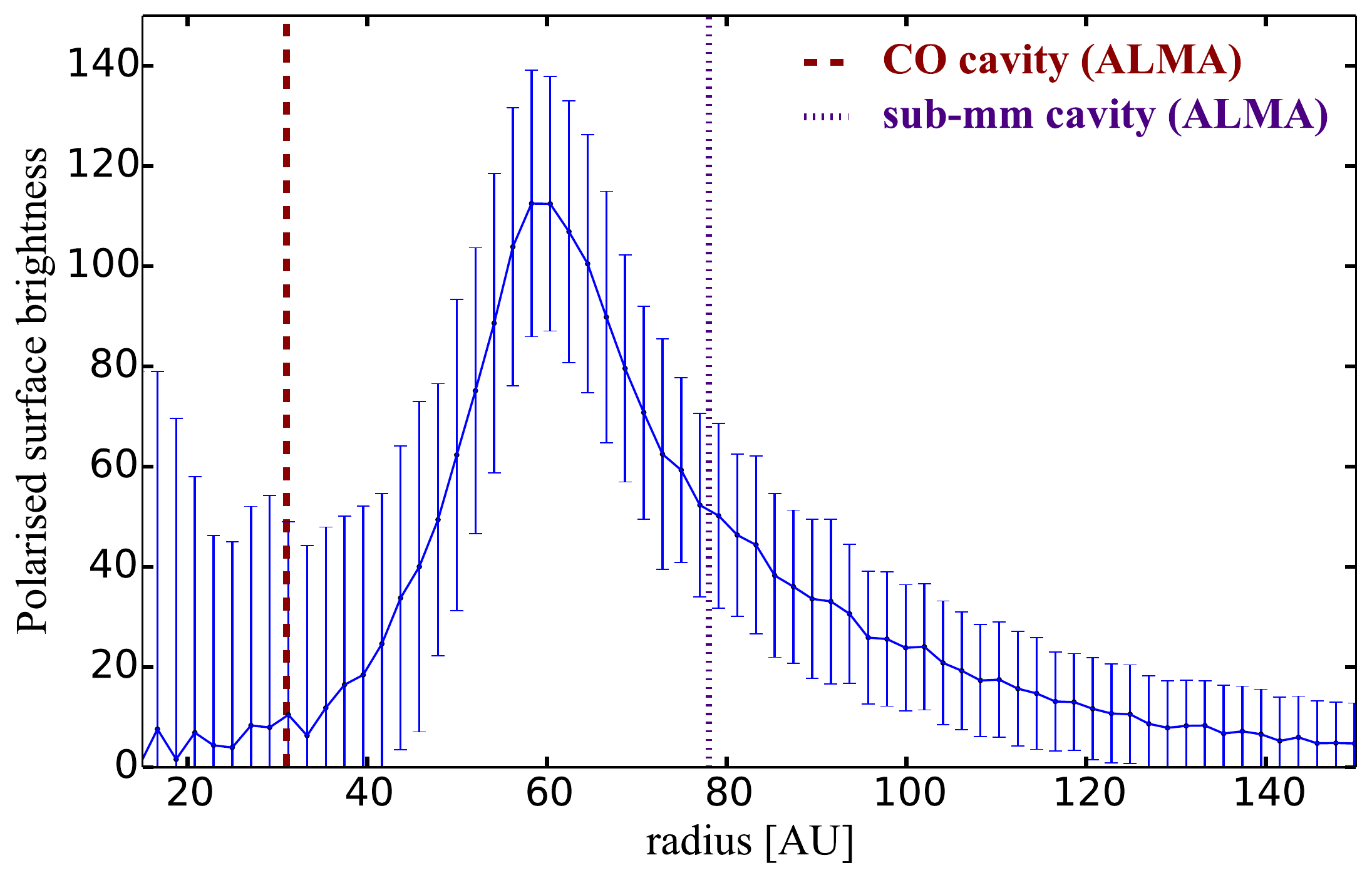}
     \end{tabular}
   \caption{{\small Left panel: overlay of the $R'$ band (0.626\,$\mu$m) $Q_\phi$ reflected light (which is \emph{not} scaled by $r^2$) and 880~$\mu$m map from ALMA Cycle~0 observations (contour lines every 10...90\% peak of the 880~$\mu$m continuum emission) of J1604. Right panel: radial profile of the polarised surface brightness (arbitrary units), and the comparison with the size of the mm-cavity observed with ALMA at 880~$\mu$m \citep{zhang2014}. The cavity radius inferred from CO $J=3-2$ emission is also displayed.  The error bars correspond to the standard deviation at each position from calculating the mean value at each radius  from the centre of the image.}\vspace{-0.2cm}}
   \label{sphere_alma_overlay}
    \end{figure*}

\section{Results} \label{sec:results}

\subsection{Radial profile}
Figure~\ref{sphere_alma_overlay} shows the overlay of the $R'$ band $Q_\phi$ reflected light and 880~$\mu$m continuum map from ALMA Cycle~0 observations  \citep[retrieved in the ALMA archive,][]{zhang2014}. The radial profile of the polarised surface brightness is also illustrated. This profile was obtained by calculating the mean value at each radius  from the centre of the $Q_\phi$ image, and the error bars correspond to the standard deviation at each position. As a result of poor seeing and moderate AO performance,  speckle noise inside a region of 0.1'' surrounding the star still dominates. Therefore, we only show the profile from the radius of confidence ($>0.1''$ corresponding to $>$15~AU at 145~pc).

The radial profile shows that  the reflected light at  0.626\,$\mu$m has a gap from 15 to 40~AU, and it has a  bright annulus  from 40 to 80~AU. The reflected light extends until ${\sim}120$~AU. We fit a Gaussian profile to the ring emission ($a\exp{[-(x-b)^2/2c^2}]+d$),  from ${\sim}40$~AU to ${\sim}80$~AU. The centre of the Gaussian ($b$) and its width ($c$) were obtained by $\chi^2$ minimisation, and the values are ${\sim}61.5\pm0.3$~AU and ${\sim}8.5\pm0.4$~AU, respectively. These findings agree with the $H$ band scattered light observations obtained with HiCIAO \citep{mayama2012}. A comparison between the HiCIAO and ZIMPOL data is shown in Appendix~\ref{appendix_a}. Compared with the ALMA observations of the $880~\mu$m continuum and CO $J=3-2$ emission, the annulus at 0.626\,$\mu$m lies inside the mm-cavity which has a radius of ${\sim}79$~AU \citep[][Fig.~\ref{sphere_alma_overlay}]{zhang2014}. The gas cavity radius was inferred around $31$~AU, but remains unresolved, which is ${\sim}$9~AU closer in than the location of the inner radius of the 0.626\,$\mu$m annulus.

The surface brightness emission beyond the peak decreases as $\propto r^{-2.92\pm0.03}$, indicating a flat  and not a flared disk \citep[a more shallow profile is expected for a flared disk, e.g.,][]{whitney1992, dalessio1998}. However, this profile is more shallow than the surface brightness profile beyond the peak from the HiCIAO data ($\propto r^{-4.70\pm0.06}$, Fig.~\ref{HiCIAO_SPHERE}). 

\subsection {Asymmetric structures}
\begin{figure*}
   \centering
   \begin{tabular}{cc}  
   \includegraphics[width=7.0cm]{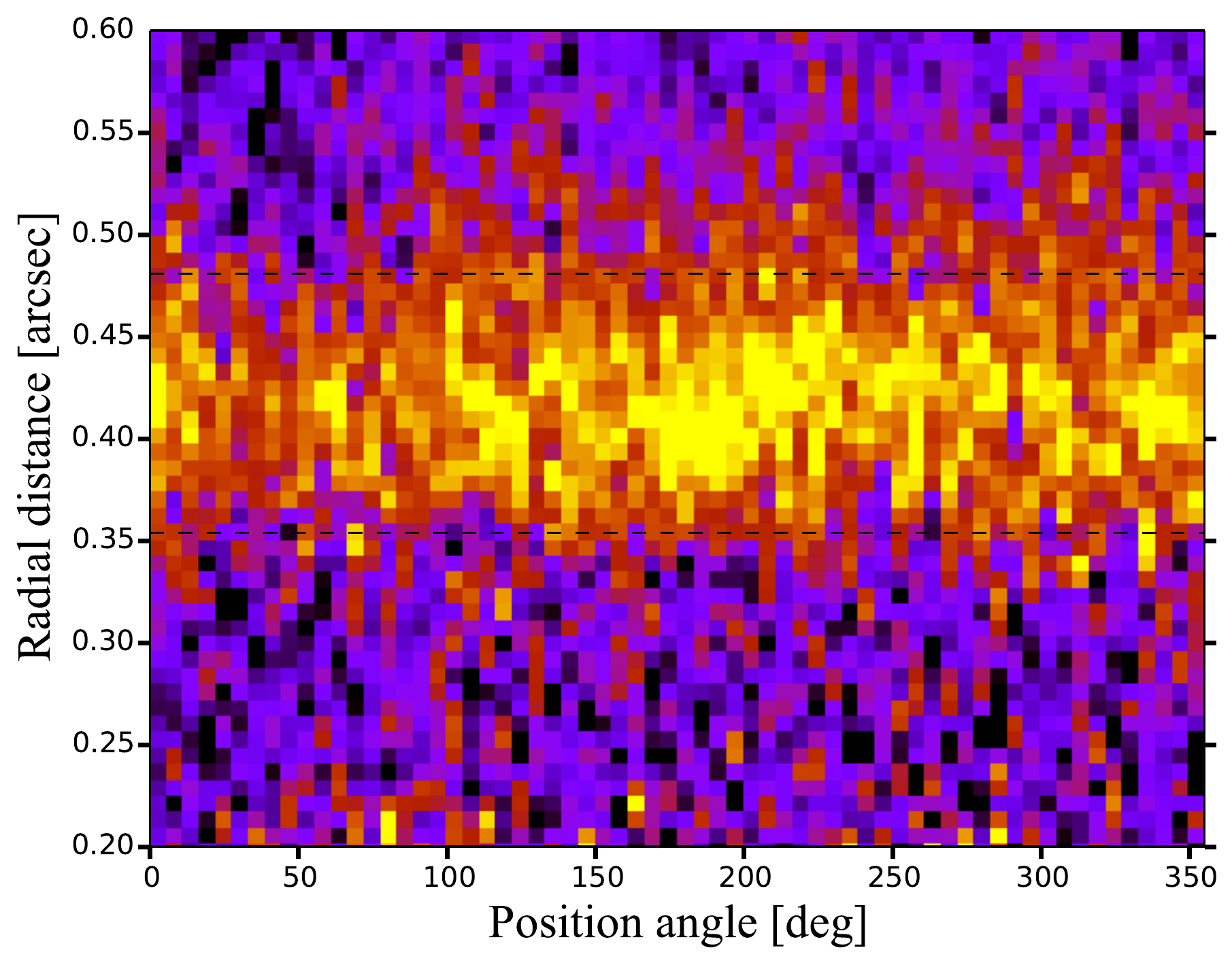}&
   \includegraphics[width=7.0cm]{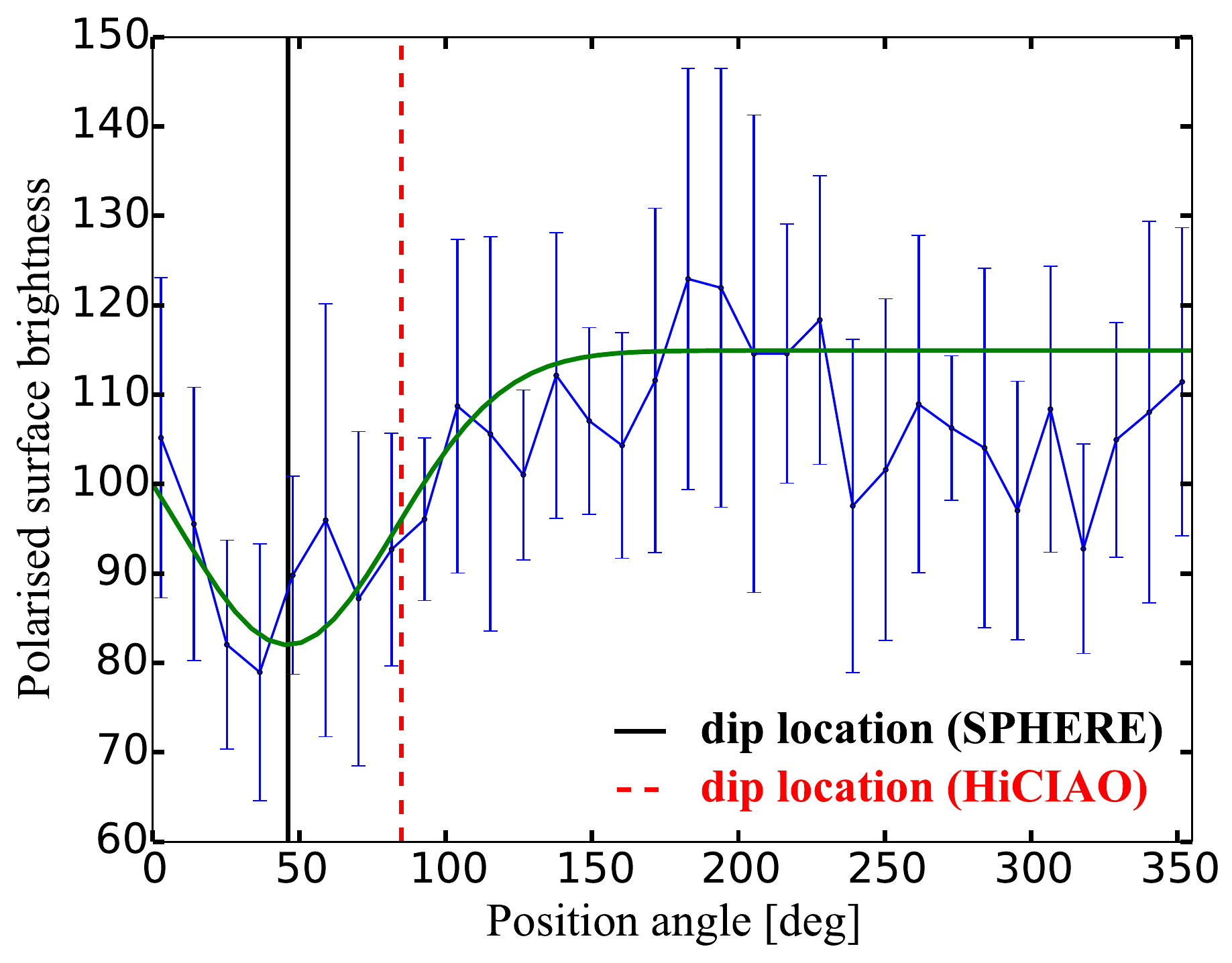}
   \end{tabular} 
      \caption{{\small Left panel: radial mapping from 0.2-0.6 arcsec of the $PI$ image of J1604 at $R'$ band. The colour scale is linear and arbitrary.  Right panel: azimuthal profile calculated from the mean values obtained between $[0.35-0.48]$~arcsec (dashed lines in the left panel and in the left panel of Fig.~\ref{sphere_images}). The error bars are from the standard deviation of the data. The solid line represents the best fit of a Gaussian by $\chi^2$ minimisation. The vertical lines show the location of the minimum of the dip from our observations and from HiCIAO observations \citep{mayama2012}.}\vspace{-0.2cm}}
         \label{azimuthal_variation}
   \end{figure*}

Figure~\ref{azimuthal_variation} shows the radial mapping from 0.2-0.6'' of the $PI$ image, which reveals one dip throughout the annulus. Since the disk is almost face-on, the map was not corrected for the inclination, because the projection would make very little difference ($\lesssim0.5\%$). An azimuthal profile  of the polarised surface brightness was obtained by taking the mean values between $0.35-0.48''$ after azimuthally binning the data by two degrees, and considering the standard deviation of the data for the error bars. The dip is clearly seen in this azimuthal profile. By fitting a Gaussian profile to the azimuthal profile (i.e. $a\exp{[-(x-b)^2/2c^2}]-d$), the best-fit parameter (by $\chi^2$ minimisation) found for the location of the dip minimum ($b$) is  ${\sim}46.2 \pm 5.4^\circ$.  Comparing the reflected light at the minimum of the dip and outside the dip, the reflected light is depleted by a factor of $\delta_{\rm{dip}}{\sim}0.72$. There are no other significant azimuthal changes of the ring morphology for different P.A. (Appendix~\ref{appendix_b}).  \cite{mayama2012} also detected a dip, but at ${\sim}85^\circ$ and with a higher contrast than our observations  $\delta_{\rm{dip}}{\sim}0.5$. We found no indication of a second dip, which was marginally detected by  \cite{mayama2012} at a P.A. of $255^\circ$. This non-detection might be due to the lower signal-to-noise of our observations.
 
The HiCIAO and the current data were taken a little more than three years apart (April 11, 2012 and 
June 10, 2015). Assuming that the dip detected with our observations is the same as was reported by \cite{mayama2012} at $85^\circ$, this would imply that the dip has a fast average rotation speed of around $12.3\pm 1.7^\circ$/year from east to north (clockwise). 

\section{Discussion} \label{sec:discussion}

The location of the edge of the gas cavity at 31~AU inferred from observations of CO $J=3-2$ of J1604 with ALMA lies inside the gap seen in scattered light (Fig.~\ref{sphere_alma_overlay}). In the context of planet disk interaction, when a  massive planet opens a gap in the disk, a spatial segregation is expected between the location of the outer edge of the gap in gas and in dust, which is predicted to become larger at longer wavelengths \citep[e.g.][]{pinilla2012}.  This is because the position of the pressure maximum at the outer edge of a gap (i.e. the location where the large particles do not experience radial drift) can be much farther out than the planet position and thus the location of the outer edge of the gap in gas. The fact that in J1604 the edge of the gas cavity lies much closer than the  inner edge of the annulus detected in our observations at $R'$ band points to a very massive planet or companion.

\cite{dejuanovelar2013} predicted the radial profile of emission at different wavelengths after performing radiative transfer together with hydrodynamical and dust-evolution modelling and combined with instrument simulations (including ZIMPOL and ALMA). A large radial segregation between the  inner edge of the annulus or ``wall'' observed with ZIMPOL polarimetric images (defined as the radial location where the flux has increased by half from the minimum in the gap and the peak of the annulus) and the peak of emission at submillimetre (850~$\mu$m) was predicted for different planet masses and locations  \citep[Fig.~8 in][]{dejuanovelar2013}. For J1604, the location of the wall of the annulus in $R'$ band is ${\sim}52$~AU, implying $r_{\rm{wall-ZIMPOL}}/r_{\rm{peak-ALMA}} {\sim}0.65$, which suggests a massive companion of $5-10~M_{\rm{Jup}}$ mass embedded around $20-40$~AU distance from the star. An upper limit of ${\sim}$18-21~$M_{\rm{Jup}}$ for a companion at 20-40~AU in J1604 has been inferred from non-redundant aperture mask interferometry \citep{kraus2008}, which does not contradict our findings. Interestingly, the location of the gas cavity inferred by observations of CO lies also in the same range.  There is a slight difference of the ring centre  between HiCIAO and SPHERE data, which is within the uncertainties of our data (approximately ${\sim}1.5$ pixel size, i.e. a shift of ${\sim}3$~AU, Fig.~\ref{HiCIAO_SPHERE}). The peak of emission at $R'$ and H band can change for different dust density distributions  in the case of planet-disk interaction \citep[shift of $\lesssim$5~AU for very massive planet $\gtrsim 15~M_{\rm{Jup}}$,][]{dejuanovelar2013}. 

If the dip in the ring of J1604 observed with HiCIAO in 2012 is the
same structure that we observe, then the dip must be rotating quickly,
with an angular speed of ${\sim}12^\circ$/year \citep[clockwise as the disk rotation derived from the CO emission][]{zhang2014}.  
The local Keplerian speed
at the position of the dip ($61$~AU) is approximately
${\sim}0.8^\circ$/year, much lower than derived from the two
observations.  Instead, the dip could be the shadow of a structure
orbiting much closer to the star.  The derived angular velocity corresponds
to a Keplerian circular orbit at a distance of only ${\sim}9.6$ AU from the star (for a $1M_\odot$ star).
The nature of the structure creating this shadow is unclear at the moment,
it could be a warp in the inner disk regions or a more localised
feature such as circumplanetary material of a planet at that
location.  If the secondary dip in the HiCIAO observations is confirmed,
an inclined inner disk might be responsible \citep{marino2015}. 
Our observations provide no direct constraints at 10 AU from
the star.

If the ring-like structure is created by a planet carving a gap, this
planet would be much closer to the ring itself (at $20-40$~AU), orbiting more slowly
than the dip.  The gap-carving planet is therefore unlikely to be
associated with the shadowing structure.

It is of course also possible that the dips seen in 2012 and
in 2015 are unrelated temporary features, or that the dip rotates in the other direction, that it has rotated over more than 360 degrees, in which cases the derived
angular velocity is meaningless and the fact that we see different
dips might be related to the observed variability of J1604. While no
mid-infrared (MIR) excess was detected in IRS spectra taken with
\textit{Spitzer }\citep{dahm2009}, photometric
data between 3 and 16~$\mu$m obtained with the Wide-field Infrared
Survey Explorer (WISE) do show MIR excess, suggesting an optically
thick narrow ring located close to the dust sublimation radius
\citep{luhman2012}. The discrepancy between WISE and IRS points to
variability of the inner disk. Rapid infrared variability has also
been detected in several other disks \citep[e.g.][]{sitko2012, flaherty2013}.

Future high-contrast observations (in a year or more from now) can
confirm whether the observed dip is the same in HiCIAO and our
observations and if it rotates with a constant speed; or if the two
observed dips are independent events, which would suggest fast inner
disk variability.

\begin{acknowledgements}
We are grateful to C.~P.~Dullemond, S.~Andrews, and A.~Kraus for their feedback and to S. Mayama for sharing the HiCIAO data. We thank the VLT team for their help during the observations. P.~P. is supported by Koninklijke Nederlandse Akademie van Wetenschappen (KNAW) professor prize to Ewine van Dishoeck. M.~B. acknowledges financial support from "Programme National de Physique Stellaire" (PNPS) of CNRS/INSU, France. A.J. acknowledges the support of the DISCSIM project, grant agreement 341137 funded by the European Research Council under ERC-2013-ADG.H.~A. acknowledges financial support from FONDECYT grant 3150643. T.~B. acknowledges support from NASA Origins of Solar Systems grant NNX12AJ04G.
\end{acknowledgements}

\bibliographystyle{aa}{\small 
\bibliography{J1604_sphere.bbl}}

\appendix

\section{Comparison with HiCIAO data} \label{appendix_a}

   \begin{figure}[h!]
   \centering
   \includegraphics[width=8.0cm]{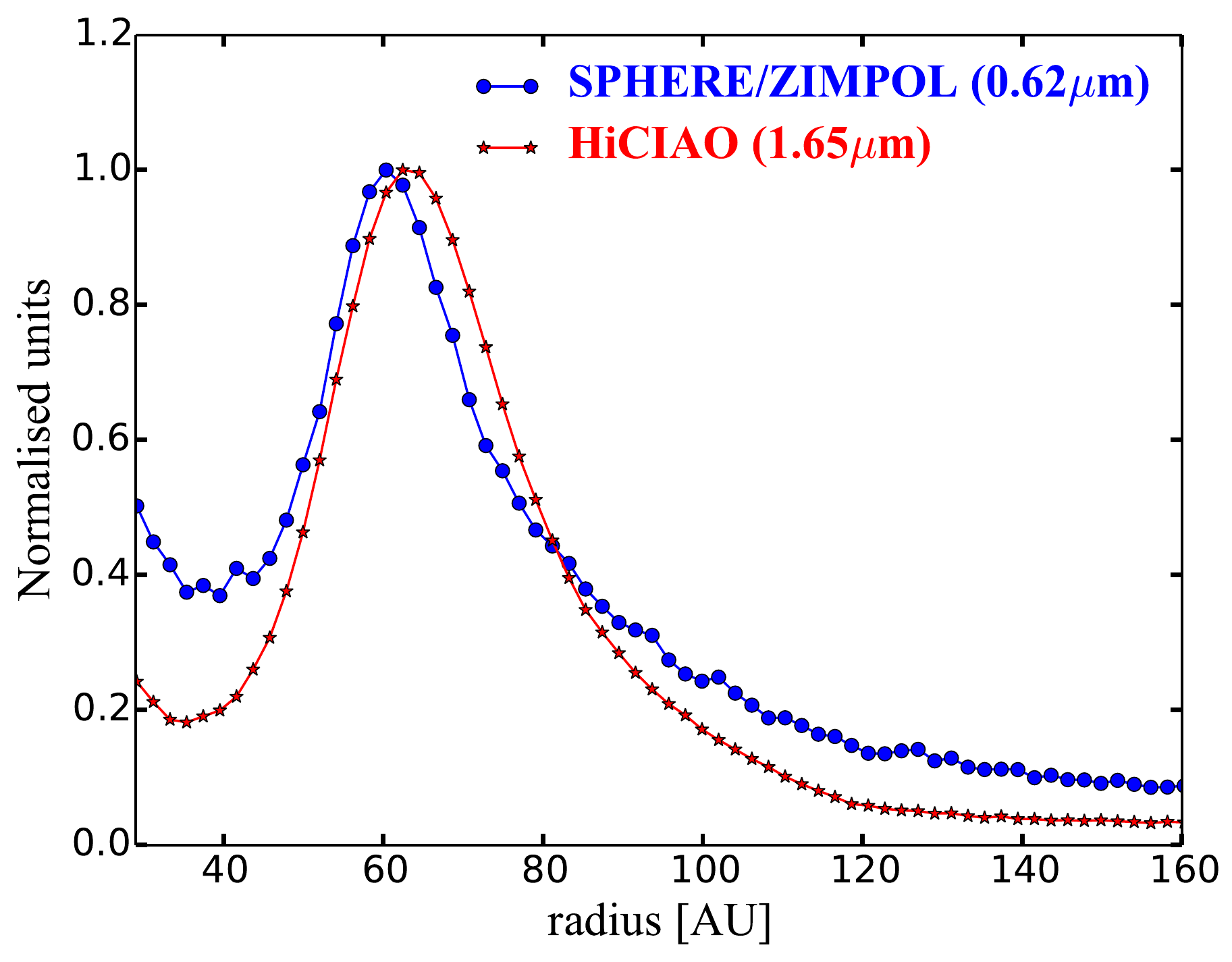}
      \caption{{\small Comparison between the azimuthally averaged radial profile of the polarised surface brightness at $R'$ and $H$ band. The data are normalised to the maximum value at $r>0.2''$. Error bars are omitted for  better readability; typical values are 40\% and 30\% of the mean value for $R'$ and $H$ band respectively.}\vspace{-0.2cm}}
         \label{HiCIAO_SPHERE}
   \end{figure}

Figure~\ref{HiCIAO_SPHERE} shows the comparison between the azimuthally averaged radial profile of the polarised surface brightness at $R'$ and $H$ band. The data are normalised to the maximum value at $r>0.2''$. By fitting a Gaussian profile to the ring emission ($a\exp{[-(x-b)^2/2c^2}]+d$), the centre of the Gaussian is at ${\sim}61.5\pm0.3$ and ${\sim}64.8\pm0.2$~AU (for 145~pc distance) for $R'$ and $H$ band, respectively. The width of the Gaussian is  ${\sim}8.5\pm0.4$ and ${\sim}10.6\pm0.9$~AU for $R'$ and $H$ band respectively. The errors are from the $\chi^2$ minimisation and are much smaller than the spatial uncertainty from the observations (1 pixel size ${\sim}$2~AU).  Fitting a power-law to the brightness profile beyond the location of the peak, the emission drops as $\propto r^{-2.92\pm0.03}$ and $\propto r^{-4.70\pm0.06}$ for $R'$ and $H$ band, respectively.

\section{Ring shape at different azimuthal cuts} \label{appendix_b}

\begin{figure}[h!]
   \centering
   \includegraphics[width=8.0cm]{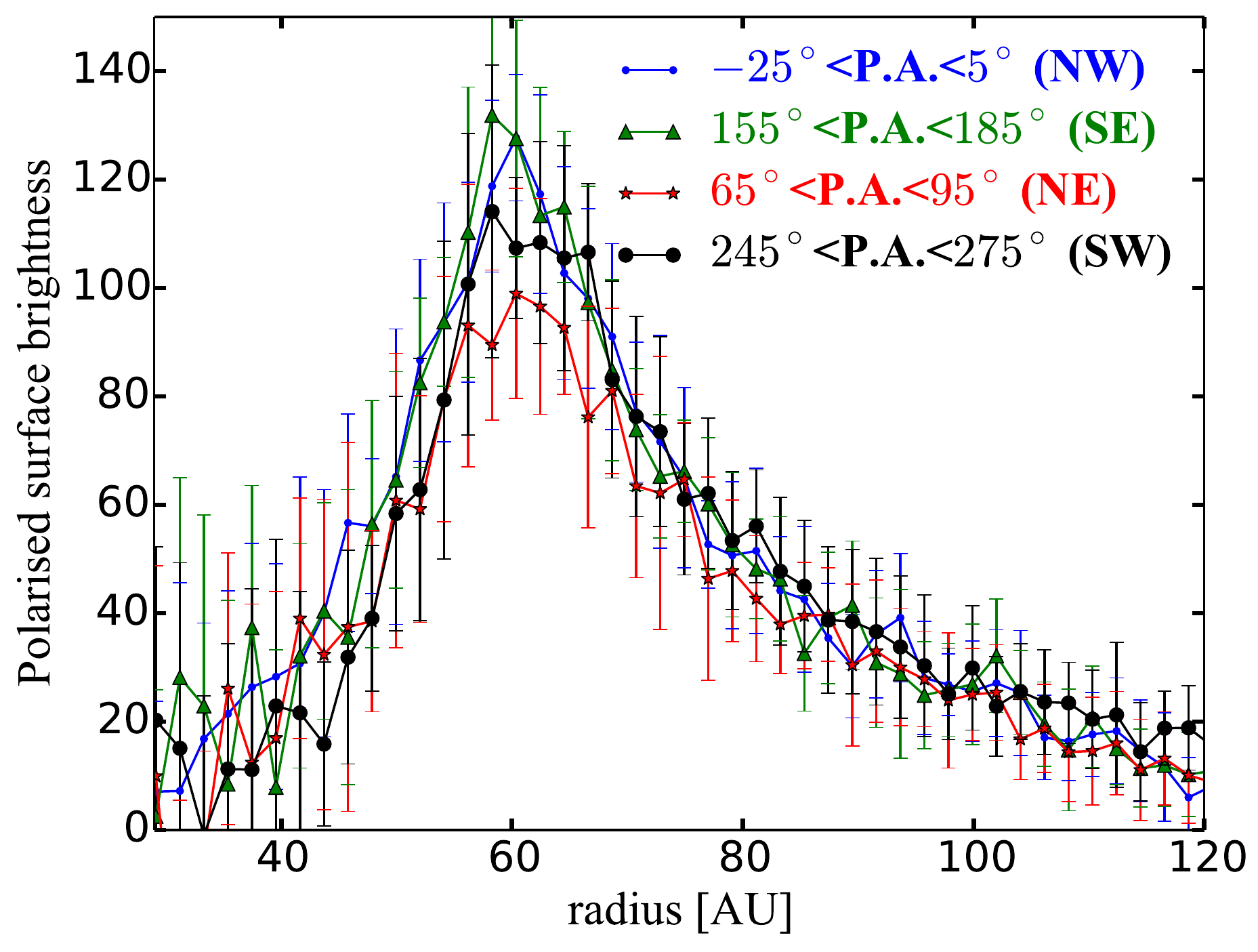}
      \caption{{\small Radial profile of the polarised surface brightness obtained assuming four different cuts of the P.A.}\vspace{-0.2cm}}
         \label{pa_cuts}
   \end{figure}

Figure~\ref{pa_cuts} shows the radial profile of the azimuthally averaged surface brightness over four bins of P.A. A Gaussian profile is fitted to each case; the width and centre of the Gaussians are summarised in Table~\ref{table_appB}. The fitting results show that there are no significant azimuthal variations of the ring within the uncertainties of the data (pixel size ${\sim}$2~AU).

%
%

\begin{table}[h!]
\caption{Centre and width of the Gaussians fits for the profiles in Fig.~\ref{pa_cuts}}
\label{table_appB}
\centering   
\tabcolsep=0.5cm          
\begin{tabular}{c|| c | c }
\hline
\hline
Cut&$b$ (centre)&$c$ (width)\\
& [AU]&[AU]\\
\hline
NW&61.4&8.7\\
\hline
SE&61.4&8.4\\
\hline
NE&62.7&9.1\\
\hline
SW&64.0&9.2\\
\hline
\end{tabular}
\tablefoot{{\small The statistical errors are omitted since they are smaller than the spatial uncertainty from the observations (1 pixel size ${\sim}$2~AU).}}
\end{table}

\end{document}